\documentclass[prd,twocolumn, superscriptaddress,floatfix,amsmath,amssymb,dvipsnames]{revtex4-1}
\usepackage{amsmath,amsfonts,amsthm,amssymb,slashed,graphicx}
\usepackage{xcolor}
\usepackage{graphicx}
\usepackage{epstopdf}
\usepackage{array,booktabs}
\usepackage{hyperref}
\usepackage{bbm} 
\usepackage{mathtools} 
\usepackage{tabulary}
\usepackage{multirow}
\usepackage{array,arydshln}
\hypersetup{linktoc = all}  
\hypersetup{pdfborderstyle={/S/U/W 0.5}}
\usepackage{tikz}
\usepackage{tkz-tab}
\usepackage{caption}
\usepackage{subcaption}
\usetikzlibrary{shapes.geometric}
\usetikzlibrary{intersections}
\usetikzlibrary{calc}
\usetikzlibrary{decorations.pathmorphing}
\usepackage[toc]{appendix}
\newcounter{mycnt}
\newcommand{\be}{\begin{equation}}
\newcommand{\ee}{\end{equation}}
\newcommand{\ba}{\begin{eqnarray}}
\newcommand{\ea}{\end{eqnarray}}

\definecolor{babyblueeyes}{rgb}{0.63, 0.79, 0.95}
\definecolor{babyblue}{rgb}{0.54, 0.81, 0.94}
\definecolor{ballblue}{rgb}{0.13, 0.67, 0.8}
\definecolor{beaublue}{rgb}{0.74, 0.83, 0.9}
\definecolor{coolblack}{rgb}{0.0, 0.18, 0.39}
\definecolor{darkcyan}{rgb}{0.0, 0.55, 0.55}
\definecolor{darkmidnightblue}{rgb}{0.0, 0.2, 0.4}
\definecolor{darkgreen}{rgb}{0.0, 0.2, 0.13}
\definecolor{upforestgreen}{rgb}{0.0, 0.27, 0.13}
\definecolor{tropicalrainforest}{rgb}{0.0, 0.46, 0.37}
\DeclareCaptionJustification{justified}{\leftskip=0pt \rightskip=0pt \parfillskip=0pt plus 1fil}

\begin{document}

\title{Topological and Time Dependence of the Action-Complexity Relation}

\author{Musema Sinamuli}
\email{cmusema@perimeterinstitute.ca}
\affiliation{Department of Physics and Astronomy, University of Waterloo, Waterloo, Ontario N2L 3G1, Canada}
\affiliation{Perimeter Institute for Theoretical Physics, Waterloo, Ontario, N2L 2Y5, Canada}

\author{Robert B. Mann}
\email{rbmann@uwaterloo.ca}
\affiliation{Department of Physics and Astronomy, University of Waterloo, Waterloo, Ontario N2L 3G1, Canada}
\affiliation{Perimeter Institute for Theoretical Physics, Waterloo, Ontario, N2L 2Y5, Canada}

\begin{abstract}
We consider the  dependence of the recently proposed action/complexity duality conjecture on time and on the underlying topology of the bulk spacetime.  For the former, we compute 
the dependence of the CFT complexity on a boundary temporal parameter and find it to be commensurate with corresponding  computations carried out in terms of  the rate of  change of the bulk action on a Wheeler deWitt (WDW) patch.  For the latter, we compare the action/complexity relation for $(d+1)$-dimensional Schwarzschild AdS black holes to those of  their geon counterparts, obtained via topological identification in the bulk spacetime. The complexity/action duality holds in both cases, but with the proportionality changed by a factor of 4, indicating sensitivity to spacetime topology.
 \end{abstract}

\maketitle

\section{Introduction}

 The importance of dualities between quantum field and gravity theories is difficult to underestimate.
The AdS/CFT correspondence \cite{thelarge}, the first and most successful, posits the existence of a 
$d$-dimensional conformal field theory (CFT) on the boundary of a $(d+1)$-dimensional asymptotically anti-de-Sitter (AdS) spacetime, and has therefore led to several dualities between quantities observed in AdS (for example black holes in the bulk) and those in the CFTs defined on their boundaries.

Recently Watanabe et.al. \cite{gravitydual} introduced a duality between a quantum information metric (or Bures metric) defined in the CFT on the boundary of an AdS black hole, and the volume of a time slice in the AdS. Their work was motivated by Susskind's idea \cite{compucomp} that it would be interesting to find a quantity in a CFT that might be  dual to a volume of a co-dimension-1 time slice of an AdS black hole spacetime.  

More recently  a similar idea was proposed suggesting a correspondence between  computational complexity in a CFT and the action evaluated on a Wheeler-De Witt (WDW) patch in the bulk \cite{complexityaction}.  In specific terms the conjecture is
\begin{equation}\label{CIeq}
C=I_{\mbox{\tiny WDW}}/\pi 
\end{equation}
where the  WDW patch refers to  the region enclosed by past and future light sheets that are sent into the bulk spacetime from a time slice on the boundary. Subsequent work \cite{gravitational,complexity} was devoted to a better understanding of how one evaluates the right-hand side of this relation. 

Complexity is concerned with quantifying the degree of difficulty of carrying out a computational task.
However a sufficiently clear  definition of  its meaning  in the CFT remains to be fully formulated.  One attempt  to this end \cite{towards} proposes a function providing a measure of the minimum number of gates necessary to reach a target state from a reference state in the CFT.  This proposal is motivated by an earlier attempt \cite{ageometric}  to provide a geometric interpretation of quantum circuits, which
consisted of the definition of two states --  a reference and a target state --  along with a unitary operator mapping the former to the latter. 
The minimum number of gates required to synthesize the unitary operator has been interpreted as a minimum length between the identity operator and that unitary operator in the manifold of unitaries. This manifold is endowed with a local metric known as the Finsler metric.  The aforementioned proposal 
\cite{towards} chose instead
the Fubini-Study metric, and the computational complexity obtained from some fixed reference and target states (related by unitaries involving a squeezing operator) appeared to be somewhat similar to the action on a WDW patch in the bulk.

Furthermore, a time dependent expression of the complexity  derived from the CFT computations remains  to be derived, despite previous work computing the rate of  change of the conjectured complexity in terms of  the rate of  change of the action on a Wheeler deWitt (WDW) patch at late time \cite{complexityaction, gravitational, complexity, onthetime}.   
 It is of particular interest to determine how  computational complexity grows in the late boundary-time limit.    Attempts to build a time-dependent complexity from CFTs \cite{comparisonof, divergences1, evolution1} yielded an expression for complexity that did not grow linearly at late time as conjectured. Furthermore, 
using a recent proposal for circuit complexity \cite{circuitcomp}, it has been shown \cite{evolution2} that
that  complexity growth dynamics has two distinct phases: an early regime whose   evolution is approximately linear is followed by a saturation phase  characterized by oscillations around a mean value.  To this end, one goal of the current paper is to compute from the CFT perspective the dependence of  complexity on boundary time  in the late time limit. 
   
The other goal of our paper is to understand if and how equation \eqref{CIeq} is sensitive to topological effects.  The  simplest spacetimes that allow the most straightforward exploration of such effects is the  
AdS black hole in  $(d+1)$ dimensions  with  an identification 
that renders it an $\mathbb{RP}^d$ geon \cite{geonwith}.   The complexity of the AdS black hole spacetimes has been studied recently \cite{complexity}, but their geon counterparts have not (though there has been recent work incorporating a different form of topological identification in the BTZ case ($d=2$) \cite{holocomplex}).  In the particular case $d=2$,  the BTZ-geon is obtained by placing further identifications on the BTZ black hole;  
the  boundary of the  Euclidean continuation of the BTZ spacetime  is an $\mathbb{RP}^2$ space, whereas that of its geon counterpart is a Klein bottle  \cite{geonwith,Friedman:1993ty}.   Previous work  \cite{Sinamuli:2016rms}   demonstrated that 
the quantum information metric \cite{gravitydual} was sensitive to spacetime topology in this case,
 and so it is reasonable to expect complexity to have a similar dual dependence on bulk topology.
   
Our paper is  organized as follows.
In section 2, the notion of complexity will be revisited and written in term of control functions, introduced as the Hamiltonian components in a basis of generalized Pauli matrices.
The same steps will be followed in section 3, but here the manifold of unitaries will be taken to be $SU(1,1)$, which is non compact. A useful expression of the complexity will then be derived.
Section 4 will specify our considerations to Gaussian states as they are very central in the understanding of quantum information processing with continuous variables. The reference and target states will both be taken to be Gaussian states.
The complexity of a $d$ dimensional CFT will be expressed in section 5, as well as its rate of change in the late time limit.  To attain this, a time-dependent target state will be chosen, and thus the unitary map between the reference and target state will have  time dependence.   
Section 6 will be devoted to the complexity of the Schwarzschild-$\mbox{AdS}_{d+1}$ spacetime and its geon counterpart as a quotient space, along with its equivalent quantum system, and  in section 7 the rate of change of the action in the bulk evaluated on a WdW patch for both the $\mbox{AdS}_{d+1}$ black hole and the $\mbox{AdS}_{d+1}$-geon will be computed. The result will be two similar correspondence relations that illustrate the sensitivity of \eqref{CIeq} to the topology of the bulk.
The last section will be a conclusion and discussion, in which our  results  will be summarized in
the context of previous work.

\section{Complexity and cost function}\label{sec2}

Here we intend to define  computational complexity in a quantum theory and study its evolution in terms of a single parameter.  We   revise the notion of   complexity introduced in \cite{ageometric} as a quantity obtained from two fixed (in time) states and a unitary operator mapping one state to the other. We follow the same steps in the case where at least one of the states (from which the complexity is constructed) is time-dependent. This complexity can be understood as  the minimum number of resources required to reach a given configuration of a quantum system starting from an initial configuration thereof. 

We will be working with quantum systems  (more specifically CFTs) whose set of unitary operators corresponds to $SU(2^n)$.   
To this end, let us consider a quantum system whose Hamiltonian in an $SU(2^n)$ basis takes the form \citep{ageometric}
\begin{eqnarray}
\label{Hamiltonian} 
H(t)=\sum_i \gamma^i(t)\sigma_i
\end{eqnarray}
where ~$\sigma_i$ are the $4^n-1$ basis matrices of $SU(2^n)$ and $\gamma^i (t)$ are the components of the Hamiltonian in that basis. These are functions of the variable $t$ defined in the interval $[s_i,s_f]$, and are referred to as  control functions.

The evolution of an arbitrary operator $V$ in the manifold $SU(2^n)$,  whose   Hamiltonian  is of the form (\ref{Hamiltonian}),  satisfies the equation \cite{ageometric}
\begin{eqnarray}
\label{unitary}
\frac{dV}{dt}=-iH(t)V~~\mbox{with}~~V(0)=I~~\mbox{and}~~V(1)=U
\end{eqnarray}
where $I$ is the identity operator. We have also defined $t$ in the interval $[s_i=0, s_f=1]$. 

We now introduce two states, an initial reference state $|R\rangle$ and a final target state $|T\rangle$,
whose relationship is given by
\begin{equation}
\label{target}
|T\rangle=U|R\rangle
\end{equation}
with ~$U$  the unitary operator introduced in (\ref{unitary}). It
  can be reached or approximated by a combination of unitary gates of $SU(2^n)$.  In this context, computational complexity is defined as an expression quantifying  the minimum number of gates or operators required to synthesize $U$.   
 
 To make this concrete  we introduce a cost function as a functional of the control function via the relation \cite{ageometric}
\begin{equation}\label{comp1}
C_f(\gamma)=\int^{1}_{0}f(\gamma(t))dt
\end{equation}
where the function~$f$~ is a given distance function.  We define  complexity by minimizing the cost function via  
\begin{equation}\label{comp2}
C_f(U)\equiv\inf_\gamma C_f(\gamma).
\end{equation}
In order to be more specific on the nature of the function $f(\gamma)$, let us define the tangent space to the unitary manifold $SU(2^n)$ at the point U as $T_USU(2^n)$ (or $T$ to be short). Thus, we identify $f(\gamma)$ with a metric function mapping elements of the tangent bundle $TM~ (M=SU(2^n))$ at a point $U$ to elements of the set of scalars  $\mathbb{R}$. That is, $f: TM\rightarrow \mathbb{R}$. 
We can reformulate   $f(\gamma)$ in terms of a new metric function via \cite{ageometric}
\begin{equation}
\label{metricfct}
F(U,y)\equiv f(\gamma)~~~~~~~~~~y\in T_U\mbox{SU}(2^n)
\end{equation}
where
\begin{eqnarray}
\label{coordinates1}
&&y=\sum_iy^i(\partial/\partial x^i)_U\nonumber\\
&&y^i=i\mbox{Tr}(\sigma_i dU/dt~U^\dagger )/2^n~~~~~~y.\sigma_i= i dU/dt~U^\dagger.
\end{eqnarray}
 The coordinates $y^i$ are determined for a given unitary operator in equation  (\ref{appendixA0}) in the appendix.

 The cost function \eqref{comp1} is proportional to the   length associated with the metric function $F(U,y)$, and will have the form \cite{ageometric} 
\begin{equation}
\label{length}
l_F(s)=\int_I dt F(s(t),[s]_t)
\end{equation}
where $s: I\rightarrow M$ maps elements of an interval $I$ to those of the manifold $M=SU(2^n)$, ~$s(t)$ is a point on the manifold and ~$[s]_t$ the tangent space to the manifold at that point.  The 
complexity measure \eqref{comp2} is obtained by minimizing $l_F(s)$ over the interval from reference to target state.

There are various different types of functions $F(U,y)$ that one can employ to 
compute (\ref{length}). We will only enumerate those that involve an $L^{(1)}$-norm and an $L^{(2)}$-norm along the path, namely \cite{ageometric}    
\begin{eqnarray}
\label{metrics}
&&F_1(U,y)\equiv\sum_i|y^i|,~~~F_p(U,y)\equiv\sum_{\sigma}p(wt(\sigma_i))|y^i|\nonumber\\
&&F_2(U,y)\equiv\sqrt{\sum_i({y^i})^2},~~F_q(U,y)\equiv\sqrt{\sum_iq(wt(\sigma_i))({y^i})^2}\nonumber\\
&&
\end{eqnarray}
where $p(wt(\sigma_i))$ and $q(wt(\sigma_i))$ are weight functions.

Suppose that the target state is a state that depends on a parameter $\sigma$ (not to be confused with the basis functions $\sigma_i$)
defined in the interval $ [s_i,s_f]$. 
The expression (\ref{target}) in this case takes the form  
\begin{equation}
\label{target2}
|\Psi(\sigma)\rangle=U(\sigma)|R\rangle.
\end{equation}
 Introducing   the Fubini-Study metric \cite{towards}  
\begin{equation}
\label{fubini}
ds_{FS}(\sigma)=d\sigma\sqrt{|\partial_\sigma|\Psi (\sigma)\rangle|^2-|\langle\Psi (\sigma)|\partial_\sigma|\Psi (\sigma)\rangle|^2}
\end{equation}
we find  
\begin{equation}
\label{length1}
l(|\Psi(s_i)\rangle,|\Psi(s_f)\rangle)=\int^{s_f}_{s_i}ds_{FS}(\sigma)
\end{equation}
yielding the length  as function of $\sigma$ associated with the FS metric.  
The above expression tells us about the evolution of the computational complexity as a function of $\sigma$. We shall postpone the question as to whether the current metric is an $L^{(1)}$-or $L^{(2)}$-norm  in the coming sections.  

\section{$SU(1,1)$ manifold and metric generation}

We now review the steps  required for the derivation of the unitary operator mapping the reference to the target state and thus the Fubini-Study metric that the unitary yields \cite{towards}, but with  complexity  reformulated to be time-dependent. For simplicity we shall deal  with quantum systems whose  manifolds of unitaries are non compact and isomorphic to $SU(2^n)$ (with $n=1$). We shall specifically work with the group $SU(1,1)$ which admits the Poincare disk as the manifold associated with its coset $SU(1,1)/U(1)$.

Coherent states, which are either characterized by complex eigenvalues of a non  compact generator of the group $SU(1,1)$ \cite{newcoherent} or by points of a coset space of the same group \cite{generalizedcoh}, can be defined for a unitary irreducible representation of $SU(1,1)$. 
$SU(1,1)$ coherent states are the result of a two mode squeezing operator
\begin{equation}
S_2(\xi)=\exp [\xi^\ast K_--\xi K_+]
\end{equation}
acting on a Fock state. ~$\xi$ is a complex parameter and $K_\pm$ are generators of the $SU(1,1)$ group that we will define explicitly in the next few steps. 
 
 We start with a target state $|\Psi (\sigma)\rangle$  (where $\sigma$ is a parameter in the time interval $[s_i,s_f]$) in a $d$ dimensional CFT, which obeys the equation (\ref{target2}) with a reference state being a two-mode state of some momentum spaces. This two-mode state consists of a product state $|\overrightarrow{k},-\overrightarrow{k}\rangle$ of two basis states, one mode representing a state of positive momentum $\overrightarrow{k}$ and the other of negative momentum   $-\overrightarrow{k}$. This can also be expressed in terms of the quantum numbers associated with the momenta ~$|n_k,n_{-k}\rangle$.  We also consider the unitary operator $U(\sigma)$ to be of the form  
\begin{eqnarray}
\label{unitary2}
U(\sigma)=e^{\int_{\Lambda}d^{d-1}k~g(\overrightarrow{k},\sigma)}
\end{eqnarray}
with
\begin{eqnarray}
g(\overrightarrow{k},\sigma)&=&\alpha_+(\overrightarrow{k},\sigma)K_+(\overrightarrow{k})+\alpha_-(\overrightarrow{k},\sigma)K_-(\overrightarrow{k})\nonumber\\
&+&\omega(\overrightarrow{k},\sigma)K_0(\overrightarrow{k})
\end{eqnarray}
 and $\Lambda$~ a momentum cut-off parameter.  Note that the direction that only gives an overall phase to the state is modded out .
 
 The quantities
$\alpha_+(\overrightarrow{k},\sigma),~\alpha_-(\overrightarrow{k},\sigma), ~\omega(\overrightarrow{k},\sigma)$ are arbitrary functions whereas  $K_+(\overrightarrow{k}),~ K_-(\overrightarrow{k})$ and $K_0(\overrightarrow{k})$ are the generators of the $SU(1,1)$ algebra. These latter quantities can be written in term of   annihilation  operators $(b_{\overrightarrow{k}},b_{-\overrightarrow{k} })$ and creation operators $(b^\dagger_{\overrightarrow{k}},b^\dagger_{-\overrightarrow{k}})$  associated with the respective modes $(\overrightarrow{k},-\overrightarrow{k})$ as \cite{towards}
\begin{eqnarray}
\label{su11}
&&K_+=\frac{1}{2}b^\dagger_{\overrightarrow{k}}b^\dagger_{-\overrightarrow{k}}\nonumber\\
&&K_-=\frac{1}{2}b_{\overrightarrow{k}}b_{-\overrightarrow{k}}\nonumber\\
&&K_0=\frac{1}{4}(b^\dagger_{\overrightarrow{k}}b_{\overrightarrow{k}}
+b_{-\overrightarrow{k}}b^\dagger_{-\overrightarrow{k}})
\end{eqnarray}
and satisfy the commutation relations
\begin{equation}
[K_+,K_-]=-K_0 ~~~~~~~[K_0,K_\pm]=\pm\frac{1}{2}K_\pm
\end{equation}

It is straightforward to show that (\ref{unitary2}) can be put into the form  \cite{agroup}
\begin{eqnarray}
\label{unitary3}
U(\sigma)&=&e^{\int_\Lambda d^{d-1}k~\gamma_+(\overrightarrow{k},\sigma)K_+(\overrightarrow{k})}\nonumber\\
&\times &
e^{\int_\Lambda d^{d-1}k~\log(\gamma_0(\overrightarrow{k},\sigma))K_0(\overrightarrow{k})}\nonumber\\
&\times &e^{\int_\Lambda d^{d-1}~k\gamma_-(\overrightarrow{k},\sigma)K_-(\overrightarrow{k})}
\end{eqnarray}
where the new functions $\gamma_+(\overrightarrow{k},\sigma), \gamma_-(\overrightarrow{k},\sigma)$ and $\gamma_0(\overrightarrow{k},\sigma)$ read as 
\begin{eqnarray}
&&\gamma_{\pm}=\frac{2\alpha_{\pm}\sinh\Xi}{2\Xi\cosh\Xi-\omega\sinh\Xi}\nonumber\\
&&\gamma_0=(\cosh\Xi-\frac{\omega}{2\Xi}\sinh\Xi)^{-2}\nonumber\\
&&\Xi^2=\frac{\omega^2}{4}-\alpha_+\alpha_-.
\end{eqnarray}

It is desirable to obtain the simplest possible form of (\ref{unitary3}).  This can be done by 
imposing  the conditions \cite{towards} 
\begin{eqnarray}
 K_-|R\rangle =0 
 \qquad  K_0|R\rangle =\frac{ \delta^{d-1} (0)}{4}|R\rangle
\end{eqnarray}
on   the reference state, yielding
\begin{eqnarray}
|\Psi (\sigma)\rangle &=&N~e^{\int_{\Lambda}d^{d-1}k~ \gamma_+(\overrightarrow{k},\sigma)K_+(\overrightarrow{k})}|R\rangle\nonumber\\
N&=& e^{\frac{1}{4}\delta^{d-1}(0)\int_\Lambda d^{d-1}k\log(\gamma_0(\overrightarrow{k},\sigma))}
\end{eqnarray}
and so  only the factor involving $\gamma_+$ needs to be taken into account. The quantity $\delta^{d-1}(0)$ comes from the commutation rules $[b_{-\overrightarrow{k}},
b^\dagger_{-\overrightarrow{k}^{'}}]=\delta^{d-1}(\overrightarrow{k}-\overrightarrow{k}^{'})$ obeyed by the operators $b_{-\overrightarrow{k}}$ that appear in the generator $K_0$.
 
Now that we have managed to find a reduced form of the unitary operator $U(\sigma)$, we will chose a reference state and attempt to derive the complexity using the Fubini-Study metric (\ref{fubini}).
 By choosing a reference state annihilated by the $b_{\overrightarrow{k}}$
\begin{equation}
|R\rangle=|0,0\rangle
\end{equation}
we obtain, when omitting the variables and the integrals
\begin{equation}
\label{zeromode}
|\Psi\rangle=Ne^{\gamma_+K_+}|0,0\rangle.
\end{equation}
We find that  (\ref{zeromode}) becomes 
\begin{equation}
\label{zeromode1}
|\Psi\rangle=\sqrt{1-|\gamma_+|^2}\sum_{n}(\gamma_+)^n|n,n\rangle
\end{equation}
upon choosing $N$ so that the target state is normalized.
Inserting (\ref{zeromode1}) in the Fubini-Study metric,
\begin{equation}
\label{complexity}
ds^2_{FS}=\langle\delta\Psi|\delta\Psi\rangle-\langle\delta\Psi|\Psi\rangle\langle\Psi|\delta\Psi\rangle
\end{equation}
we get  (see also appendix (\ref{appendix0}))
\begin{equation}
\label{metric2}
ds^2_{FS}=\frac{|\delta\gamma_+|^2}{(1-|\gamma_+|^2)^2}.
\end{equation}
Restoring the variables and the integrals, we obtain a more general form of the complexity (\ref{length1}) with the expression 
\begin{eqnarray}
\label{complexity1}
C^{(n)}&=&\min_{\gamma_+}\int^{s_f}_{s_i}d\sigma\sqrt[n]{\frac{V_{d-1}}{2}\int d^{d-1}k~|ds_{FS}(\sigma)/d\sigma|^n}\nonumber\\
&&
\end{eqnarray}
with ~$\gamma_+^{'}=\partial\gamma_+/\partial\sigma$ and $V_{d-1}$  the $(d-1)$-dimensional volume of a time slice.  Upon comparison with \eqref{metrics} we see that   (\ref{complexity1}) is an $L^{(n)}$-norm. 

We will mostly use the case where $n=1$  
\begin{eqnarray}
\label{complexity2}
C^{(1)}&=&\min_{\gamma_+}\int^{s_f}_{s_i}d\sigma ~\frac{V_{d-1}}{2}\int d^{d-1}k\frac{|\gamma_+^{'}|}{1-|\gamma_+|^2} \nonumber\\
&&
\end{eqnarray}
as it leads to a function easier to integrate as well as to a complexity whose   rate of change corresponds to that of the action evaluated in the bulk.  Note that the gates for different k's are not allowed to act in parallel in order to obtain the $C^{(1)}$ norm.


\section{Gaussian states}

 Here we briefly review the Gaussian states of a quantum system  \cite{towards}.  Such  states play a central role in quantum information processing with continuous variables as well as in quantum field theory where the vacuum states of some field theories (for example, quantum electrodynamics)
 appear to be Gaussian states.  We shall choose  the reference and target states
 to be Gaussian states.

Consider a scalar field theory in a $d$ dimensional spacetime with the Hamiltonian density 
\begin{equation}
\label{hamiltonian0}
H_m=\frac{1}{2}\int d^{d-1}x~[\pi^2+(\partial_x \Phi)^2+m^2\Phi^2] 
\end{equation}
where $m$ is the mass of the field $\Phi (x)$ and $\pi (x)$ is its conjugate momentum. These obey the commutation rules
\begin{equation}
\label{commutation}
[\Phi (\overrightarrow{x}),\pi (\overrightarrow {x}^{'})]=i\delta^{d-1}(\overrightarrow{x}-\overrightarrow{x}^{'}).
\end{equation}
The field and its conjugate momentum in terms of the annihilation $a_k$ and creation operators $a^\dagger_k$ are explicitly given by 
\begin{eqnarray}
\label{scalar1}
\Phi (x)&=&\int d^{d-1}k\frac{1}{\sqrt{2\omega_k}}(a_k~e^{-ikx}+a^\dagger_k~e^{ikx})\nonumber\\
\pi (x)&=&\int d^{d-1}k\frac{\sqrt{\omega_k}}{\sqrt{2}i}(a_k~e^{-ikx}-a^\dagger_k~e^{ikx})
\end{eqnarray}
with ~ $\omega_k=\sqrt{k^2+m^2}$.
Substituting   (\ref{scalar1}) into  (\ref{commutation}) we find
\begin{equation}
[a_{\overrightarrow{k}},a^\dagger_{\overrightarrow{k}^{'}}]=\delta^{d-1}(\overrightarrow{k}-\overrightarrow{k^{'}})
\end{equation}
with all other commutators zero. 

It is helpful to write things in   momentum space where the Hamiltonian can be expressed in a more elegant form as
\begin{equation}
\label{hamiltonian1}
H_m=\int d^{d-1}k~\omega_k~ \big[a^\dagger_{\overrightarrow{k}}a_{\overrightarrow{k}}+\frac{1}{2}\big] 
\end{equation}
 and the field and its associated momentum become
\begin{eqnarray}
\label{scalarfield}
&&\Phi(\overrightarrow{k})=\frac{1}{\sqrt{2\omega_k}}(a_{\overrightarrow{k}}+a^\dagger_{-\overrightarrow{k}})\nonumber\\
&&\pi(\overrightarrow{k})={\frac{\sqrt{\omega_k}}{\sqrt{2}i}}(a_{\overrightarrow{k}}-a^\dagger_{-\overrightarrow{k}}).
\end{eqnarray} 
In the sequel we consider a CFT  for which the field is massless   $(m=0)$.
 
 A pure Gaussian state $|S\rangle$ is a state  for which \cite{towards}
\begin{equation}\label{GS}
\big[\sqrt{\frac{\alpha_k}{2}}\Phi(\overrightarrow{k})+\frac{i}{\sqrt{2\alpha_k}}\pi(\overrightarrow{k})\big]|S\rangle =0
\end{equation}
where ~$\alpha_k=\omega_k$ corresponds to the ground state $|m\rangle$ ~of the theory . We can consider the target state to be the ground state. 

To construct the reference state $|R(M)\rangle$ we write the Bogoliubov transformation \cite{towards}
\begin{eqnarray}
\label{bogoliubov}
b_{\overrightarrow{k}}=\beta^+_k a_{\overrightarrow{k}}+\beta^-_k a^\dagger_{-\overrightarrow{k}}
\end{eqnarray}
and require 
\begin{equation}\label{refcond}
b_{\overrightarrow{k}}|R(M)\rangle=0.
\end{equation}
where  $\beta^+_k=\cosh{2r_k}~,~~\beta^-_k=\sinh{2r_k} $ and $r_k=\log (\sqrt[4]{M/\omega_k})$.  This corresponds to a state with~$\alpha_k=M$ in \eqref{GS}. 

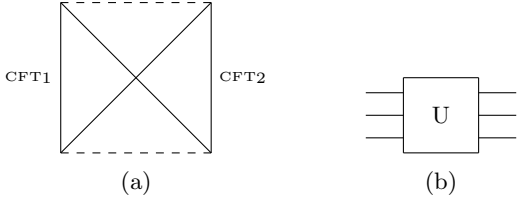
\begin{figure}
\begin{subfigure}[b]{0.22\textwidth}
\centering
\begin{tikzpicture}[scale=1]
\draw[dashed](0,0)--(2,0);
\draw(2,0)--(2,2)
node[midway, right, inner sep=1mm] {$\mbox{{\tiny CFT}}_2$};
\draw[dashed](2,2)--(0,2);
\draw(0,2)--(0,0)
node[midway, left, inner sep=1mm] {$\mbox{{\tiny CFT}}_1$};
\draw(0,0)--(2,2);
\draw (2,0)--(0,2);
\end{tikzpicture}
\caption{}\label{fig:M1}
\end{subfigure}
\begin{subfigure}[b]{0.22\textwidth}
\centering
\begin{tikzpicture}[scale=1]
\draw (0.5,0.5)--(1.5,0.5);
\draw (0.5,1.5)--(1.5,1.5);
\draw (0.5,0.5)--(0.5,1.5);
\draw (1.5,0.5)--(1.5,1.5);
\draw (0,1)--(0.5,1);
\draw (0,0.7)--(0.5,0.7);
\draw (1.5,0.7)--(2,0.7);
\draw (0,1.3)--(0.5,1.3);
\draw (1.5,1.3)--(2,1.3);
\draw (1.5,1)--(2,1);
\node at (1,1)   (a) {U};
\end{tikzpicture}
\caption{}\label{fig:M2}
\end{subfigure}
\caption{ (a) Conformal diagram of a BTZ ($d=2$) black hole. As we can see, a CFT is defined at each boundary thereof.
(b) Quantum circuit which consists in an unitary $U$ acting on $n$ qubits. In the context of the current work its associated complexity can be regarded as equivalent to the action integral evaluated on a WDW patch in BTZ black hole.}\label{fig:M3}
\end{figure}
 
\section{Conformal Field Theory in $d$ dimensions}

Employing the formalism of the previous sections, we now compute the complexity defined in the CFT dual of an AdS gravitational theory.  The spacetimes we have in mind for the latter are  AdS black holes which, according to the AdS/CFT correspondence, admit CFTs on their boundaries.  The Penrose diagram for the  $\mbox{AdS}_{d+1}$ black hole is illustrated in figure \ref{fig:M3}. The BTZ case can be described as a quotient space of $\mbox{AdS}_{d+1}~~\mbox{with}~~d=2$.  

Here we aim to derive the computational complexity associated to quantum theories defined in the boundary CFTs.    States on such CFTs  are described  by thermofield double (TFD) of finite temperature, defined in a thermal circle of period $\beta$ 
\cite{eternalblack} 
\begin{align}
\label{tfd1}
|\mbox{TFD(t)}\rangle &\equiv e^{-i(H_1+H_2)t}|\mbox{TFD(0)}\rangle \nonumber \\
& = e^{-i(H_1+H_2)t}\sum_{n}e^{-\beta E_n/2}|n\rangle_1|n\rangle_2
\end{align}
 with $H_{1,2}$ the free Hamiltonians, ~$|n\rangle_{1,2}$ the eigenstates of the free Hamiltonians defined on the $\mbox{CFT}_{1,2}$ and $E_n$ their corresponding energies.  These states on the $\mbox{CFT}_1$ can be assigned to the positive momentum modes $\overrightarrow{k}$ and the ones on the $\mbox{CFT}_2$ to the negative momentum modes $-\overrightarrow{k}$ of a scalar field theory.
 
We see that
\begin{eqnarray}
\label{tfd0}
|\mbox{TFD(0)}\rangle&\equiv&\sum_{n}e^{-\beta E_n/2}|n\rangle_1|n\rangle_2\nonumber\\
&=&e^{\int d^{d-1}k~ e^{-\beta\omega_k/2}a^\dagger_{\overrightarrow{k}}
a^\dagger_{-\overrightarrow{k}}}|0\rangle
\end{eqnarray}
for a free scalar field theory. The state $|TFD(0)\rangle$ is annihilated by operators $b_{\pm\overrightarrow{k}}$ defined via a Bogoliubov transformation as
\begin{eqnarray}
\label{bogo}
b_{\overrightarrow{k}}&=&\cosh\theta_k a_{\overrightarrow{k}}-\sinh\theta_k a^\dagger_{-\overrightarrow{k}}\nonumber\\
b_{-\overrightarrow{k}}&=&\cosh\theta_k a_{-\overrightarrow{k}}-\sinh\theta_k a^\dagger_{\overrightarrow{k}}
\end{eqnarray}
with~ $\tanh\theta_k=e^{-\beta\omega_k/2}$.
 
\vskip 5pt We can regard the states in the boundaries as two-mode states where one side of the diagram (figure \ref{fig:M1}) corresponds to states of a conformal scalar field theory with positive momentum $\overrightarrow{k}$ and the other side to a scalar field theory with negative momentum states $-\overrightarrow{k}$. The total Hamiltonian of the system according to (\ref{hamiltonian1}) will be     
\begin{eqnarray}
\label{totalham}
H&=&H_1+H_2\nonumber\\
&=&\int d^{d-1}k~\omega_k[a^\dagger_1a_1+a^\dagger_2a_2+1]
\end{eqnarray}
where ~$\omega_k=k$, ~$a_1=a_{\overrightarrow{k}}~~\mbox{and}~~a_2=a_{-\overrightarrow{k}}$.
Using \eqref{bogo}, 
the total Hamiltonian (\ref{totalham}) in the basis   (\ref{su11})  has the form  
\begin{equation}
a^\dagger_1a_1+a^\dagger_2a_2+1=4\cosh(2\theta_k)K_0+2\sinh(2\theta_k)(K_++K_-)
\end{equation}
and so  (\ref{tfd1}) becomes  
\begin{equation}
\label{tfd2}
|\mbox{TFD}\rangle\equiv e^{\alpha_+ K_++\alpha_-K_-+\omega K_0}~|TFD(0)\rangle
\end{equation}
with  
\begin{eqnarray}
&&\alpha_{\pm}=-2i~\omega_k~t\sinh(2\theta_k)\nonumber\\
&&\omega=-4i~\omega_k~t\cosh(2\theta_k).
\end{eqnarray}
Equation (\ref{tfd2}) will become
\begin{equation}
\label{tfd4}
|\mbox{TFD}\rangle\equiv e^{\gamma_+K_+}e^{\log(\gamma_0)K_0} e^{\gamma_-K_-}~|TFD(0)\rangle 
\end{equation}   
 using the transformation of the unitary operator (\ref{unitary3}).

 We obtain
a state equivalent to (\ref{zeromode}) and (\ref{zeromode1}), but 
where 
\begin{eqnarray}
&&\gamma_{\pm}=\frac{-i\sinh(2\theta_k)\sin\Xi}{\cos\Xi+i\cosh(2\theta_k)\sin\Xi}\nonumber\\
&&\Xi=2\omega_k~t~~~~\mbox{and}~~~~\omega_k=k.
\end{eqnarray}
  In term of the parameter $\sigma$ the control function $\gamma_+$ can be written as
\begin{eqnarray}
&&\gamma_\pm(k,\sigma)=\frac{-i\sinh(2\theta_k)\sin\Xi}{\cos\Xi+i\cosh(2\theta_k)\sin\Xi}\nonumber\\
&&\Xi=2kt~\sigma.
\end{eqnarray}
 It is easy to check that $\gamma_+=\gamma_+(k,\sigma)$ as a function of $\sigma$, satisfies the conditions
\begin{eqnarray}
\gamma_+(k,s_i)&=&0~~~~\mbox{and}\nonumber\\
\gamma_+(k,s_f)&=&\frac{-i\sinh(2\theta_k)\sin (2kt)}{\cos (2kt)+i\cosh(2\theta_k)\sin (2kt)}
\end{eqnarray}
corresponding to reference and target state respectively. It appears that the control function is time-dependent and this fact will imply a time-dependent complexity.
  
 In order to compute the complexity in the simplest possible manner we consider situations in which the control function obeys the condition $|\gamma_+|<1$, which is holds if the operator is unitary.

Now that we have assembled all the ingredients, the complexity (\ref{complexity2}) as a function of $t$ is  
\begin{eqnarray}
C^{(1)}(t)&=&\min_{\gamma_+}\int^{s_f}_{s_i}d\sigma ~\frac{V_{d-1}}{2}\int d^{d-1}k~\frac{|\gamma_+^{'}|}{1-|\gamma_+|^2}\nonumber\\
&=&2V_{d-1}\Omega_{\kappa,d-2}\beta^{-d}(2^d-1)\Gamma (d)\zeta (d)~t
\label{complexC1}
\end{eqnarray}
as detailed in eq.  (\ref{appendix1}) in the appendix.  The computational complexity can be understood as the minimum number of gates needed to synthesize a unitary operator $U$ (figure \ref{fig:M2}). 

Before proceeding further, we define the total energy of the scalar field as (see \eqref{scalarenergy} in the appendix)
\begin{eqnarray}
E&=&V_{d-1}\int d^{d-1}k~ \omega_k e^{-\beta \omega_k}\nonumber\\
&=&V_{d-1}\Omega_{d-2}\beta^{-d}\Gamma (d).
\end{eqnarray}
Hence the complexity \eqref{complexC1} takes the form 
\begin{equation}
\label{adscomplex}
C^{(1)}(t)= 2(2^d-1)\zeta (d) E~ t.
\end{equation}
 Note that the rate of change of the complexity for  very large $t$ is
\begin{eqnarray}
\label{rated2}
{\frac{dC(t)}{dt}}^{\mbox{\tiny AdS}_{d+1}}=n_d E
\end{eqnarray}
with $n_d=2(2^d-1)\zeta (d)$ a dimensionless constant.  Equation (\ref{rated2}) means that the variation of the complexity with respect to time at late time is proportional to the  total energy $E$ of the CFT . This total energy $E$ will later be identified with the mass of the AdS black hole dual to the CFT.

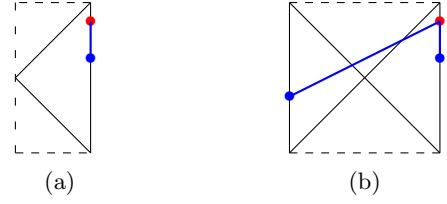
\begin{figure}
\begin{subfigure}[b]{0.22\textwidth}
\centering
\begin{tikzpicture}[scale=1]
\draw[dashed](1,0)--(2,0);
\draw(2,0)--(2,2);
\draw[dashed](2,2)--(1,2);
\draw[loosely dashed](1,2)--(1,0);
\draw(1,1)--(2,2);
\draw (1,1)--(2,0);
\node [red] at (2,1.75) {\textbullet};
\node [blue] at (2,1.25) {\textbullet};
\draw[thick,blue](2,1.75)--(2,1.25);
\end{tikzpicture}
\caption{}\label{fig:M0a}
\end{subfigure}
\begin{subfigure}[b]{0.22\textwidth}
\centering
\begin{tikzpicture}[scale=1]
\draw[dashed](0,0)--(2,0);
\draw(2,0)--(2,2);
\draw[dashed](2,2)--(0,2);
\draw(0,2)--(0,0);
\draw(0,0)--(2,2);
\draw (2,0)--(0,2);
\node [red] at (2,1.75) {\textbullet};
\node [blue] at (2,1.25) {\textbullet};
\node [blue] at (0,0.75) {\textbullet};
\draw[thick,blue](2,1.75)--(2,1.25);
\draw[thick,blue](2,1.75)--(0,0.75);
\end{tikzpicture}
\caption{}\label{fig:M0b}
\end{subfigure}
\caption{(a) The current diagram shows the thermofield single state on the boundary of the BTZ-geon. The red and blue points represent the right- and left-modes of the thermofield single (both modes or CFTs are superposed on the same boundary), respectively. The thick blue line corresponds to the complexity derived from the entangled state.
(b) This diagram corresponds to the first one but here the situation is seen in the BTZ context. Unfolding the CFTs on the first diagram, the left-modes appear on both sides of the new diagram (which a BTZ one). It results a sum of two complexities. }
\label{fig:M0c}
\end{figure}

\section{Geon and Direct Products}

In this section we repeat these  computations in the context of the  $\mbox{AdS}_{d+1}$-geon.  

The   $\mbox{AdS}_{d+1}$ black hole has the metric  
\begin{eqnarray}
\label{adsd+1}
ds^2&=&-f(r)dt^2+dr^2/f(r)+r^2d\Sigma^2_{\kappa,d-1}\nonumber\\
f(r)&=&\kappa -\omega^{d-2}/r^{d-2}+r^2/l^2
\end{eqnarray}
which, in Kruskal coordinates $(\tilde{U},\tilde{V}, x^i)~~\mbox{with}~~i=1~\mbox{to}~d-1$,~~ takes the form
\begin{equation}
ds^2=-fd\tilde{U} d\tilde{V}+r^2 d\Sigma^2_{\kappa,d-1}(x^i)
\end{equation}
where
$f~\mbox{and}~ r$ are smooth functions of $(\tilde{U},\tilde{V})$.

The $\mbox{AdS}_{d+1}$-geon is the quotient spacetime resulting from a freely activing involutive isometry applied to the $\mbox{AdS}_{d+1}$ black hole \cite{geonwith}.
It is obtained via the identification \cite{geonwith, looking}
\begin{equation}
J:(\tilde{U},\tilde{V},x^i)\rightarrow (\tilde{V},\tilde{U},P(x^i))
\end{equation}
which corresponds to the change 
\begin{equation}
\label{identification}
(t,x^i)\rightarrow (-t,-x^i)
\end{equation}
in the spacetime coordinates. $P(x^i)=-x^i$ is the antipodal map on the $(d-1)$-dimensional sphere $S^{d-1}$,~ which corresponds to $\kappa=1$ in (\ref{adsd+1}).

  The state associated with the CFT on the geon boundary is the thermofield single \cite{behind}
\begin{equation}
|\Psi_g\rangle=e^{-(\beta/4+it)H}|C\rangle
\end{equation}
where  $|C\rangle$ is the cross-cap state, consisting of an entangled state between left- and right- moving modes of a free boson CFT (see figure \ref{fig:M0a}). In terms of the modes $j_n~\mbox{and}~ \bar{j}_n$ of the holomorphic and anti-holomorphic conserverd currents $J=i\partial X~\mbox{and}~ \bar{J}=i\bar{\partial}X$, respectively, it is solution to \cite{introtoconf}
\begin{equation}
[j_n+(-1)^n\bar{j}_{-n}]|C\rangle=0 
\end{equation}
and thus takes the form
\begin{equation}
|C\rangle=\exp\bigg[-\sum^{\infty}_{n=1}\frac{(-1)^n}{n}j_{-n}\bar{j}_{-n}\bigg]|0\rangle
\end{equation} 
which clearly shows entanglement between the left- and right-moving modes of the CFT.

In the case of the geon space, we claim that due to the reflection coming from the involution $J$ the metric function $F(U,y)$, satisfies
\begin{equation}
\label{geonmetric}
F(U,y)_{\tiny{\mbox{Geon}}}\leq F(U, y)_{\tiny{\mbox{BTZ}}}+F(U^{'}, y^{'})_{\tiny{\mbox{BTZ}}}.
\end{equation}
The right-hand side of (\ref{geonmetric}) saturates the geon metric function. This make sense when the complexity is regarded as the minimum time required to approximate the unitary. The presence of first and second terms on the right hand side of (\ref{geonmetric}) is depicted in figures \ref{fig:M0a} and \ref{fig:M0b}.

\begin{figure}
\begin{subfigure}[b]{0.22\textwidth}
\centering
\begin{tikzpicture}[scale=1]
\draw[dashed](1,0)--(2,0);
\draw(2,0)--(2,2)
node[midway, right, inner sep=1mm] {$\mbox{{\tiny CFT}}_1\equiv \mbox{{\tiny CFT}}_2$};
\draw[dashed](2,2)--(1,2);
\draw[loosely dashed](1,2)--(1,0);
\draw(1,1)--(2,2);
\draw (1,1)--(2,0);
\end{tikzpicture}
\caption{}\label{fig:M4}
\end{subfigure}
\begin{subfigure}[b]{0.22\textwidth}
\centering
\begin{tikzpicture}[scale=1]
\draw (0.5,0)--(1.5,0);
\draw (0.5,1)--(1.5,1);
\draw (0,0.5)--(0.5,0.5);
\draw (1.5,0.5)--(2,0.5);
\draw (0.5,0)--(0.5,1);
\draw (1.5,0)--(1.5,1);
\draw (0,1.7)--(0.5,1.7);
\draw (1.5,1.7)--(2,1.7);
\draw (0,2.3)--(0.5,2.3);
\draw (1.5,2.3)--(2,2.3);
\node at (1,2)   (a) {U};
\draw (0.5,1.5)--(1.5,1.5);
\draw (0.5,2.5)--(1.5,2.5);
\draw (0,2)--(0.5,2);
\draw (1.5,2)--(2,2);
\draw (0.5,1.5)--(0.5,2.5);
\draw (1.5,1.5)--(1.5,2.5);
\draw (0,0.8)--(0.5,0.8);
\draw (1.5,0.8)--(2,0.8);
\draw (0,0.2)--(0.5,0.2);
\draw (1.5,0.2)--(2,0.2);
\node at (1,0.5)   (a) {V};
\end{tikzpicture}
\caption{}\label{fig:M5}
\end{subfigure}
\caption{(a) Conformal diagram of a BTZ geon. The two CFTs, one at each boundary are now identified in only one boundary.
(b) Quantum circuit composed of unitaries $U$ acting on $n_A$ qubits and $V$ acting on $n_B$ qubits (when $V=I$, the $n_B$ qubits are ancilla ones). This circuit complexity corresponds to the action integral evaluated on a WDW patch in the BTZ geon space. }
\label{fig:M6}
\end{figure}
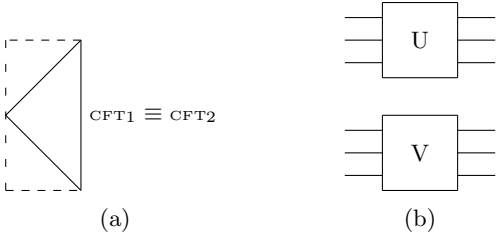

Thus  the unitary operator $U^{'}$ and the tangent space vectors $y^{'}$ to the manifold of unitary operators at $U^{'}$ correspond to those where the spacetime coordinates for the left-modes are  $(-t,-x^i)$.
Equation (\ref{geonmetric}) can be understood as the metric function of a quantum system  consisting of the direct product of two other  quantum systems (figures \ref{fig:M4} and  \ref{fig:M5}). Indeed, let us suppose that $F_A,~F_B$, and $F_{AB}$ are the metrics  given in equation (\ref{metricfct}) on $SU(2)^{n_A},~SU(2)^{n_B}$ and $SU(2)^{n_A+n_B}$, respectively.      The metric $F_{AB}$ of the system composed of a unitary $U$ on the $n_A$ qubit and a unitary $V$ on the $n_B$ qubits is \cite{ageometric} 
\begin{equation}
\label{additive}
F^2_{AB}(U\otimes V, H_A+H_B)=F^2_A(U, H_A)+ F^2_B(V, H_B)
\end{equation}
where $H_A\in SU(2)^{n_A}$ and $H_B\in SU(2)^{n_B}$ (omitting the tensor factors $I_A\otimes .$~ and~ $.\otimes I_B$ acting trivially on $V$ and $U$, respectively).  The Finsler metrics $F_A,~F_B$ and $F_{AB}$ are said to form an {\it additive triple} of Finsler metrics. Equation (\ref{additive}) leads to the inequality
\begin{equation}
F_{AB}(U\otimes V, H_A+H_B)\leq F_A(U, H_A)+ F_B(V, H_B).
\end{equation}
.

 The quantity we are now going to compute is the complexity corresponding to the metric $F(U^{'},y^{'})$ in (\ref{geonmetric}).  We first introduce the notion of an F-Isometry.
 A map $h: s(t)\rightarrow h(s(t))$ is an F-Isometry if and only if the length (\ref{length}) associated with the metric $F(s[t], [s]_t)$ satisfies the relation 
\begin{eqnarray}
l_F(s)=l_F(h~o~s)
\end{eqnarray}
and 
\begin{equation}
F(s[t], [s]_t)=F((h~o~s)(t), [h~o~s]_t).
\end{equation}
In the tangent space to the manifold at $s(t)$, it acts like
\begin{equation}
[h~o~s]_t=h_\ast[s]_t
\end{equation}
with $h_\ast$ defined as 
\begin{eqnarray}
&&h_\ast:T_{s(t)}M\rightarrow T_{h(s(t))}M
\end{eqnarray}
such that the F-Isometry reads as 
\begin{equation}
F(x,y)=F(h(x),h_{\ast}y).
\end{equation}
Under the identification (\ref{identification}), the momentum components transform as  
\begin{eqnarray}
&& k_0\equiv\frac{\partial}{\partial t}\rightarrow \frac{\partial}{\partial (-t)}=-\frac{\partial}{\partial t}\equiv-k_0\nonumber\\
&& k_i\equiv\frac{\partial}{\partial x^i}\rightarrow\frac{\partial}{\partial (-x^i)}=-\frac{\partial}{\partial x^i}\equiv -k_i
\end{eqnarray}
From the above relations we infer that the quantities~$k=\sqrt{\sum^{d-1}_{i=1}k^2_i}$,~~~and~~~$\Xi=2\omega_k t$~~with~~$(\omega_k\rightarrow -\omega_k,~t\rightarrow -t)$ are invariant under these transformations.
 Hence  the control function  
\begin{eqnarray}
\gamma_+=\frac{-i\sinh(2\theta_k)\sin (2kt)}{\cos (2kt)+i\cosh(2\theta_k)\sin (2kt)}
\end{eqnarray}
is still invariant under these transformations.  Thus, the geon transformation is an F-Isometry,
 and still obeys the condition  $|\gamma_+|<1$.   

The complexity is therefore equal to twice that of the $\mbox{AdS}_{d+1}$ black hole since the two contributions from the geon metric contribute equally to the complexity 
\begin{eqnarray}
\label{btzgeonA}
C^{(1)}(t)&=&\min_{\gamma_+}\int^{s_f}_{s_i}d\sigma ~V_{d-1}\int d^{d-1}k~\frac{|\gamma_+^{'}|}{1-|\gamma_+|^2}\nonumber\\
&=&2 n_d E~t
\end{eqnarray}
and the rate of change thereof is
\begin{equation}
\label{btzgeonB}
{\frac{dC(t)}{dt}}^{\mbox{\tiny Geon}}=2 n_d E.
\end{equation}
 Equations (\ref{btzgeonA}) and (\ref{btzgeonB}) hold for any  $(d+1)$ dimensional $\mbox{AdS}$ geon with $d\geq 2$.

 For any  limiting value of $t$, the geon complexity is still twice the amount obtained in (\ref{adscomplex}).  More explicitly, we have  
\begin{equation}
\label{complexitygeon}
C^{\mbox{\tiny Geon}}(t)=2~C^{\mbox{\tiny AdS}_{d+1}}(t).
\end{equation}

\section{Rate of variation of the action}

 In this section we verify the  action-complexity conjecture in the context in which we have been working:   between an action evaluated in the bulk (on a particular patch) and the complexity computed in the CFTs at the boundaries of the Schwarzschild AdS  black holes and their geon counterparts.

\begin{figure}
\begin{subfigure}[b]{0.22\textwidth}
\centering
\begin{tikzpicture}[scale=1]
\draw[dashed](0,0)--(2,0);
\draw(2,0)--(2,2);
\draw[dashed](2,2)--(0,2);
\draw(0,2)--(0,0);
\draw[thick,tropicalrainforest](0.9,2)--(0,1.1);
\draw[thick,tropicalrainforest](1.1,2)--(2,1.1);
\draw[thick,tropicalrainforest](0,0.9)--(0.9,0);
\draw[thick,tropicalrainforest](1.1,0)--(2,0.9);
\node [red] at (0.9,2) {\textbullet};
\node [red] at (0,1.1) {\textbullet};
\node [red] at (1.1,2) {\textbullet};
\node [red] at (2,1.1) {\textbullet};
\node [red] at (0,0.9) {\textbullet};
\node [red] at (0.9,0) {\textbullet};
\node [red] at (1.1,0) {\textbullet};
\node [red] at (2,0.9) {\textbullet};
\draw[fill=babyblueeyes!30](0.9,2)--(0,1.1)--(0,0.9)--(0.9,0)--(1.1,0)--(2,0.9)--(2,1.1)--(1.1,2)--cycle;
\end{tikzpicture}
\caption{}\label{fig:M7}
\end{subfigure}
\begin{subfigure}[b]{0.22\textwidth}
\centering
\begin{tikzpicture}[scale=1]
\draw[dashed](1,0)--(2,0);
\draw(2,0)--(2,2);
\draw[dashed](1,2)--(2,2);
\draw (1,0)--(1,2);
\draw[thick,tropicalrainforest](1.1,2)--(2,1.1);
\draw[thick,tropicalrainforest](1.1,0)--(2,0.9);
\node [red] at (1.1,2) {\textbullet};
\node [red] at (2,1.1) {\textbullet};
\node [red] at (1.1,0) {\textbullet};
\node [red] at (2,0.9) {\textbullet};
\draw[fill=babyblueeyes!30](1,2)--(1,0)--(1.1,0)--(2,0.9)--(2,1.1)--(1.1,2)--cycle;
\end{tikzpicture}
\caption{}\label{fig:M8}
\end{subfigure}
\caption{
(a) Conformal diagram of a BTZ black hole with its WDW patch. The coloured area in light blue is the area over which we evaluated the bulk contribution. The green lines are the null hypersurfaces and the red points are the joints that involve null hypersurfaces with spacelike and timelike ones.
(b) Conformal diagram of the geon space with the WDW patch on it. It is obvious to notice that only half of the coloured area, the green lines and red points in the BTZ diagram appear for the geon space.
}
\label{fig:M9}
\end{figure}
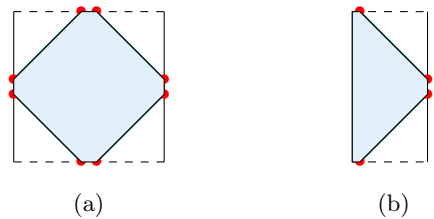

Consider a Schwarzschild-AdS black hole in $d+1$ dimensions whose metric is given by
\begin{eqnarray}
ds^2&=&-fdt^2+dr^2/f+r^2d\Sigma^2_{k,d-1}\nonumber\\
f&=&\frac{r^2}{l^2}+k-\frac{\omega^{d-2}}{r^{d-2}}
\end{eqnarray}
where ~$k=0$~ for planar black holes.
 We aim to compute the action evaluated on a WDW patch, as shown in the figure \ref{fig:M7}, for this black hole. The different contributions to the action from the bulk and the boundary terms are \cite{complexity, gravitational} 
\begin{eqnarray}
\label{wdwaction}
I&=&\frac{1}{16\pi G_N}\int_M d^{d+1}x\sqrt{-g}\big(R+\frac{d(d-1)}{l^2}\big)\nonumber\\
&+&\frac{1}{8\pi G_N}\int_B d^d\sqrt{h}K
-\frac{1}{8\pi G_N}\int_{B^{'}}d\lambda d^{d-1}\theta\sqrt{\gamma}~\kappa\nonumber\\ &+&\frac{1}{8\pi G_N}\int_\Sigma
d^{d-1}x\sqrt{\sigma}\eta
+\frac{1}{8\pi G_N}\int_{\Sigma^{'}}d^{d-1}x\sqrt{\sigma}a\nonumber\\
&&
\end{eqnarray}
with the cosmological constant (not to be confused with the cut-off parameter in the CFTs)~ $\Lambda=-d(d-1)/(2l^2)$ and the curvature radius $R=-d(d+1)/l^2$ . 

 The first term in (\ref{wdwaction}) accounts for the bulk contribution. The other terms are the boundary contributions. The second term is the surface or Gibbons-Hawking-York term, in which $K$ represents the extrinsic curvature. The third term comes from the null hypersurfaces with $\kappa$ a parameter related to the tangent vector to these hypersurfaces. The fourth term (Hayward term) is a joint term involving the junctions of spacelike/timelike hypersurfaces \cite{gravitaction, quasilocal, moving,Booth:2001gx}. The last term is also a joint term involving the junctions of null hypersurfaces.  
  
 Evaluating the bulk contributions, we obtain for the four quadrants of figure \ref{fig:M7}
\begin{eqnarray}
I_{\tiny Bulk}&=&\frac{1}{16\pi G_N}\int_M d^{d+1}x\sqrt{-g}\big(R+\frac{d(d-1)}{l^2}\big)\nonumber\\
&=&\frac{\Omega_{k,d-1}~d}{8\pi G_N l^2}\int^{r_{{\tiny max}}}_{0}dr~ r^{d-1}(v_\infty -r^\ast (r))
\end{eqnarray}
where~ $v=t+r^\ast$~ and ~ $r^\ast=\int dr/f$.
 The surface contributions lead, for the four quadrants in figure \ref{fig:M7}, to 
\begin{eqnarray}
I_{{\tiny GHY}}&=&\frac{1}{8\pi G_N}\int_B d^d x\sqrt{|h|}K\nonumber\\
&=&\frac{\Omega_{k,d-1}~d~\omega^{d-2}}{16\pi G_N}(v_\infty -r^\ast (0))
\end{eqnarray}
with ~$h$~ the induced metric on the surface. The only nonzero contributions are those coming from the singularities ($r=0$). 

 The null surface contributions are
\begin{eqnarray}
I_{{\tiny Null}}=-\frac{1}{8\pi G_N}\int_{B^{'}}d\lambda d^{d-1}\theta\sqrt{\gamma}~\kappa 
\end{eqnarray}
with ~$x^\mu=(\lambda, \theta^A)$~ parametrizing the null hypersurfaces and $\gamma$ the induced metric on them. $\kappa$ satisfies the equation $k^\mu\nabla_\mu k_\nu=\kappa k_\nu$~ and~ $k^\mu=\frac{\partial x^\mu}{\partial\lambda}$ are the tangent vectors to these surfaces. It is possible to choose everything to be affinely parametrized such that $\kappa=0$.  We thus can infer that the null surfaces do not contribute to the action.
 The joint term (Hayward) contributions have the form
\begin{eqnarray}
I_{{\tiny Hay}}=\frac{1}{8\pi G_N}\int_\Sigma
d^{d-1}x\sqrt{\sigma}\eta.
\end{eqnarray}
In our case there is no contribution coming from this term since there are no spacelike/timelike junctions for the chosen patch (figure \ref{fig:M7}).
 The contribution of the last term for the four quadrants is
\begin{eqnarray}
I_{{\tiny jnt}}&=&\frac{1}{8\pi G_N}\int_{\Sigma^{'}}d^{d-1}x\sqrt{\sigma}a\nonumber\\
&=&\frac{\Omega_{k,d-1}}{16\pi G_N}\epsilon_0^{d-1}\log (\epsilon_0^{d-2}/\omega^{d-2}).
\end{eqnarray}
It is important to recall that here the only non zero contributions are those of the junctions at the region near the singularities ($r=\epsilon_0$ with  $\epsilon_0$ very small). And we also have to keep in mind that those contributions only appear when we consider black holes with hyperbolic metrics ($k=-1$) whose horizon radii are smaller than the AdS radius ($r_h<l$).  We shall not consider these kinds of black holes any further; they lead to similar conclusions.
  
 After summing up all these contributions we find that the rate of change of the action at late time is
\begin{eqnarray}
\label{actionbtz}
\frac{dI}{dt}\bigg|_{t\rightarrow\infty}&=&\frac{1}{\pi}\frac{d}{dt}\big[I_{{\tiny Bulk}}+I_{{\tiny GHY}}\big]\bigg|_{t\rightarrow\infty}\nonumber\\
\frac{dI}{dt}\bigg|_{t\rightarrow\infty}&=&2M_\ast
\end{eqnarray}
with ~$M_\ast$~ given in appendix (\ref{massterm}).  We shall see in the next few steps that the mass term $M_\ast$ can be identified with  the total energy $E$ of the scalar field.

 Focusing now on the geon case, since in figure \ref{fig:M8} only half of the patch (two quadrants) contributes to the action, it implies that the total action for the geon space will be the half of that of the  $\mbox{AdS}_{d+1}$ black hole. 

 In fact, the time in the geon conformal diagram (see figure \ref{fig:M8}) is moving up for both the left and right CFTs. The geon action can be interpreted in the AdS context as
\begin{equation}
\label{geonaction}
I_{Geon}(t_1+t_2)=I_{{\tiny AdS}}(t_1+t_2)+I_{{\tiny AdS}}(t_1-t_2).
\end{equation}
 This can be justified by the fact that a given point in the geon diagram has two images in the AdS diagram. For symmetric time evolution ($t_1=t_2=t/2$) the second term of the right-hand side of \eqref{geonaction} is time independent whereas the first term is time dependent and is only evaluated on half the patch of the AdS black hole.
 
\vskip 5pt The rate of change at late time for the geon action then becomes 
\begin{equation}
\frac{dI}{dt}\bigg|_{t\rightarrow\infty}=M_\ast
\end{equation}
We thus obtain for  $d\geq2$ the relation
\begin{equation}
\label{actiongeon}
I^{\mbox{\tiny Geon}}=\frac{1}{2}~ I^{\mbox{\tiny AdS}_{d+1}}.
\end{equation}
Setting the \ total energy $E$ of the CFTs to be equal to the mass term $M_\ast$ of the  $\mbox{AdS}_{d+1}$ black hole, we infer that the complexity (\ref{rated2}) defined in the CFTs at the boundaries of the  $\mbox{AdS}_{d+1}$  black holes can be expressed in term of the $\mbox{AdS}_{d+1}$ action (\ref{actionbtz}) as follows  
\begin{equation}
\label{complexitybtz1}
C^{\mbox{\tiny AdS}_{d+1}}=\frac{n_d}{2}I^{\mbox{\tiny AdS}_{d+1}}.
\end{equation}
Equation \eqref{complexitybtz1} is the conjectured relation.

Making use of the equations (\ref{complexitygeon}) and (\ref{actiongeon}) we find the same relation for the $\mbox{AdS}_{d+1}$ geon  
\begin{equation}\label{geonCompAct}
C^{\mbox{\tiny Geon}}=2n_d I^{\mbox{\tiny Geon}}
\end{equation}
except for a factor of 4, indicative of the sensitive of complexity to the underlying topology of the spacetime.

 In \cite{holocomplex} the action was computed at $t=0$ for the BTZ-geon on a WDW patch partitioned into non-intersecting pieces associated with each boundary and a remaining interior piece. It was found that the action evaluated on each partition is precisely half the WDW patch-action of the corresponding two-sided BTZ wormhole ($t=0$) and is independent of the black hole mass.

\section{Conclusion}

 We have derived the computational complexity of a CFT defined on the boundary of an  $\mbox{AdS}_{d+1}$ black hole as a function of a temporal variable $t$,
  and have explicitly computed the  small-$t$ and large-$t$ limits.  The quantity $t$  
  can be regarded as the boundary time parameter, yielding the rate of change of the CFT complexity. Up to a factor this equals  $n_d$ times the rate of change of the bulk action evaluated on a WDW patch as conjectured \cite{gravitational, complexityaction}.

Our results are commensurate with 
previous work \cite{towards}, where the target state was   defined for a fixed value of  time and  where a different control function was employed, resulting in  a dimensionless complexity proportional to $V_{d-1}\Lambda^{d-1}$  (see discussions in the appendices). 
Similar results have been derived in the context of the cMERA circuit \cite{entanglementrenorm, holographicgeom1, holographicgeom2}.
 
In contrast to this, we began with a  particular configuration of the TFD state defined on the boundaries of an  $\mbox{AdS}_{d+1}$  black hole as the target state 
and obtained a more complex control function depending on the parameter  $t$. This  led us to a dimensionless expression \eqref{complexC1} for the complexity that is a function of  $t$, which is proportional to $V_{d-1}\Lambda^{d-1}$ as well  (see appendix E). 

We have also established a correspondence between the geon quotient space of the  $\mbox{AdS}_{d+1}$ black hole and a quantum system consisting of a product of two quantum systems.  We found that the complexity of the CFT on the boundary of the $\mbox{AdS}_{d+1}$ geon is twice that of the its $\mbox{AdS}_{d+1}$ black hole counterpart.  Furthermore, we found  that the rate of change of the bulk action of the  $\mbox{AdS}_{d+1}$ geon evaluated on a WDW patch is half of that of the  $\mbox{AdS}_{d+1}$ black hole. 
 
 We therefore infer that the complexity/action relationship is sensitive to the topology of the bulk spacetime: 
 there exists the same kind of correspondence relation between the  complexity of a CFT and the bulk action of a geon evaluated on a WDW patch (\ref{geonCompAct}), but with the additional (topological) factor of 4.
 
 It would be interesting to compute in future investigations the computational complexities $C^{(n)}~ (\mbox{with}~n>1)$ associated with the same control function $\gamma_+(\overrightarrow{k},\sigma)$ and see whether they can lead to desired and more general  forms of the complexity $C^{(1)}$. Likewise an exploration of   the computational complexities $C^{(n)}~ (\mbox{with} ~n\geq 1)$   for charged and/or rotating AdS black holes (and their geon counterparts \cite{geonwith}) should also provide further insight.  
  
\section*{Appendix}
\subsection{Coordinates on the tangent plane}
\renewcommand{\theequation}{A-\arabic{equation}}
\setcounter{equation}{0}

 Consider the manifold of unitaries $SU(2^n)$ and the unitary operator \cite{ageometric}
\begin{equation}
U=\exp [-i\sum_j\gamma^j\sigma_j]
\end{equation}
thereof, the tangent to $SU(2^n)$  at this point $U$, admits the coordinates 
\begin{eqnarray}
\label{appendixA0}
y^i&=&i\mbox{Tr}(\sigma_i dU/dt~U^\dagger )/2^n =d\gamma^i/dt.
\end{eqnarray}
For the metric function $F_1(U,y)=f(\gamma)$, the complexity or length (Euclidean distance) associated with it reads
\begin{eqnarray}
C_f(U)&=&\inf_{\gamma}\int_I f(\gamma (t))~dt 
= \inf_\gamma\int_I \sum_i d\gamma^i.
\end{eqnarray}
In the Poincare disk model (with $\gamma^i ~(i=+)$), the complexity or length (hyperbolic distance) associated with the metric $F_1(U,y)$ has the form 
\begin{eqnarray}
\label{appendix0}
C_f(U)=\inf_{\gamma_+}\int\frac{d\gamma_+}{1-|\gamma_+|^2}.
\end{eqnarray}

\subsection{Complexity}
\renewcommand{\theequation}{B-\arabic{equation}}
\setcounter{equation}{0}

This subsection is devoted to the derivation of the final form of the computational complexity $C^{(1)}(t)$. As introduced earlier in the previous sections, it has the form 
\begin{eqnarray}
\label{appendix1}
C^{(1)}(t)&=&\min_{\gamma_+}\int^{s_f}_{s_i}d\sigma~\frac{V_{d-1}}{2}\int d^{d-1}k~\frac{|\gamma_+^{'}|}{1-|\gamma_+|^2}\nonumber\\
&=&\int^{s_f}_{s_i}d\sigma~\frac{V_{d-1}}{2}\int d^{d-1}k~|2\omega_k t\sinh(2\theta_k)|\nonumber\\
&=&2V_{d-1}~t~\Omega_{\kappa, d-2}\int k^{d-1}\frac{e^{-\beta k/2}}{1-e^{-\beta k}}dk\nonumber\\
&=&2V_{d-1}\Omega_{\kappa,d-2}(2^d-1)\beta^{-d}\Gamma (d)\zeta (d)~t 
\end{eqnarray}
where we employ the control function
\begin{equation}
\label{appendix2}
\gamma_+=\frac{-i\sinh(2\theta_k)\sin (2kt\sigma)}{\cos (2kt)+i\cosh(2\theta_k)\sin (2kt\sigma)}
\end{equation}
yielding in turn
\begin{equation}
\label{appendix3}
\frac{|\gamma_+^{'}|}{1-|\gamma_+|^2}=2\omega_k t\sinh(2\theta_k)
\end{equation}
with 
\begin{equation}
\sinh(2\theta_k)
=\frac{2e^{-\beta\omega_k/2}}{1-e^{-\beta\omega_k}}.
\end{equation} 

\subsection{AdS/CFT (Planar black holes)}
\renewcommand{\theequation}{C-\arabic{equation}}
\setcounter{equation}{0}

Here we review some useful notions on the metric of Schwarzschild-AdS black hole, particularly the planar one, as well as the metric of its boundary CFT.
 
A planar Schwarzschild-AdS black hole in $d+1$ dimension  has the metric
\begin{eqnarray}
\label{appendix4}
ds^2&=&-fdt^2+dr^2/f+r^2d\Sigma^2_{\kappa,d-1}\nonumber\\
f&=&-\omega^{d-2}/r^{d-2}+r^2/l^2.
\end{eqnarray}
Changing variables to $z=l/r$, (\ref{appendix4}) becomes
\begin{eqnarray}
ds^2&=&\frac{l^2}{z^2}[-hd\tilde{t}^2+dz^2/h+d\Sigma^2_{\kappa,d-1}]\nonumber\\
h&=&1-(z/z_0)^d
\end{eqnarray}
where ~$\tilde{t}=t/l,~R=l,~z_0^d=l^{d-2}/\omega^{d-2}$~ and ~$\omega^{d-2}=r_h^d/l^2$.
 
 The mass  of this black hole is
\begin{equation}
\label{massterm}
M_\ast=\frac{d-1}{16\pi G_N}\Omega_{0,d-1}\omega^{d-2}
\end{equation}
The metric of the CFT on the boundary of the black hole is of the form  
\begin{equation}
ds^2_{\tiny {boundary}}=-dt^2+l^2d\Sigma^2_{\kappa,d-1}
\end{equation}
($\kappa=0$)~ for planar black holes. It can be rewritten as
\begin{equation}
\label{boundarycft}
ds^2_{\tiny {boundary}}=-l^2[d\tilde{t}^2+d\Sigma^2_{\kappa,d-1}]
\end{equation}
and we can label $\tilde{t}$ as $t$. 
 
\subsection{Total energy of the scalar field}
\renewcommand{\theequation}{D-\arabic{equation}}
\setcounter{equation}{0} 

 Here we compute the total energy of the scalar field knowing the probability densities of the Hamiltonian eigenstates $|n,n\rangle$. \vskip 5pt Starting with the state $|TFD(0)\rangle$ in \eqref{tfd0} we find that the density matrix is obtained from the expression
\begin{eqnarray}
\rho&=&Tr(|TFD(0)\rangle\langle TFD(0)|)\nonumber\\
&=&\sum_{n_k} e^{-\beta\omega_k}|n_k\rangle\langle n_k|
\end{eqnarray}
after tracing over the states $|n_k\rangle_2$,
where ~$e^{-\beta\omega_k}$ are clearly the probability densities of the Hamiltonian eigenstates. From the above expression we infer that the total energy of the scalar field reads as
\begin{eqnarray}
\label{scalarenergy}
E&=&V_{d-1}\int d^{d-1}k~ \omega_k~ e^{-\beta\omega_k}\nonumber\\
&=&V_{d-1}\int d^{d-1}k~ k~ e^{-\beta k}\nonumber\\
&=&V_{d-1}\Omega_{\kappa,d-2}\beta^{-d}\Gamma(d).
\end{eqnarray}
. 

\subsection{Comparing methods for Computing Complexity}
\label{AppE}
\renewcommand{\theequation}{E-\arabic{equation}}
\setcounter{equation}{0} 

We compare here our approach in section \ref{sec2} to a recent proposal  \cite{circuitcomp} in which a lattice was used to study the complexity of a free scalar field theory. The distinction between the two approaches consists of  the choice of gates, the distance or metric function,  and the regularization method.

1. {\it Choice of Gates}  The approach of ref. \cite{circuitcomp} is to minimize over all   gates obtained by considering the exponential of bilinear generators of the form $\Phi(x_1)\pi(x_2)$ (squeezing operator). They found that optimal circuits (in absence of penalty factors in the cost functions) admit normal mode decompositions and require for their construction only generators of the form $\Phi(\overrightarrow{k})\pi(-\overrightarrow{k})+
\pi(\overrightarrow{k})\Phi(-\overrightarrow{k})$, which are momentum preserving.
These generators have the form $G_k=\tilde{x}_k\tilde{p}_{-k}+\tilde{p}_k\tilde{x}_{-k}$  with $\tilde{x}_k=\frac{1}{\sqrt{N}}\sum^{N-1}_{a=0}\exp(-\frac{2\pi ik a}{N})x_a$ on the lattice.

By contrast, in our approach we consider Hamiltonian operators  consisting of combinations of generators $G_{2k}=\Phi(\overrightarrow{k})
\Phi(-\overrightarrow{k})$  and $G_{3k}=\pi(\overrightarrow{k})
\pi(-\overrightarrow{k})$. We thus minimize over the gates constructed from the generators $G_{2k}$ and $G_{3k}$.

 2. {\it Choice of Metric}  Instead of a Finsler metric \cite{circuitcomp} (as studied by Nielsen \cite{ageometric}),  
we use the Fubini-Study metric, and subsequently derive a time-dependent complexity which reads as
\begin{equation}
C^{(1)}\sim V_{d-1}\beta^{-d} t=V_{d-1}\beta^{-(d-1)}(t/\beta)
\end{equation}
with $\beta$ the period of the thermal circle in which is defined the TFD state (reference state).

3. {\it Choice of Regularization Method} 
The methods of  \cite{circuitcomp} yielded the result
\begin{equation}
C^{(2)}=\frac{1}{2}\sqrt{\sum^{N-1}_{k_i=0}(\log\frac{\tilde{\omega}_k}{\omega_0})^2}
\end{equation}
for the complexity \eqref{complexity1} with $n=2$, 
where the frequencies
\begin{equation}
{\tilde{\omega}_k}^2=m^2+\frac{4}{\delta^2}\sum^{d-1}_{i=1}\sin^2\frac{\pi k_i}{N}\nonumber
\end{equation}
and where $\delta$ is the lattice spacing. In $d-1$ dimensions, the lattice volume  is $V=L^{d-1}=N\delta^{d-1}$ with $N$ the number of sites. For QFTs the complexity is dominated by ultraviolet (UV) modes ($\tilde{\omega}_k=1/\delta$). The leading term thus reads as
\begin{equation}
\label{jeffmyers1}
C^{(2)}\sim \big(\frac{V}{\delta^{d-1}}\big)^{1/2}=N^{1/2}
\end{equation}
The square root in \eqref{jeffmyers1} comes from the cost function $F_2$.

To obtain an expression similar to the one proposed in \cite{comments1}:
\begin{equation}
\label{jeffmyers3}
C_{\tiny hol}\sim\frac{V}{\delta^{d-1}}=N
\end{equation}
an $F_1$ cost function was employed \cite{circuitcomp}, 
yielding the complexity
\begin{equation}
C^{(1)}\sim\frac{V}{\delta^{d-1}}\log(\frac{1}{\omega_0\delta})
\end{equation}
where $\omega_0$ is some arbitrary frequency.

  A similar result  \cite{towards} was obtained by considering the same set of gates, i.e. $G_{1k}=\Phi(\overrightarrow{k})
\pi(-\overrightarrow{k})+\pi(\overrightarrow{k})
\Phi(-\overrightarrow{k})$ employed in ref. \cite{circuitcomp} along with a Fubini-Study metric.  The
 complexity was found to have the form
\begin{equation}
C^{(n)}\sim V_{d-1}^{\frac{1}{n}}\Lambda^{\frac{d-1}{n}}\log(M/\Lambda)
\end{equation}
which, when $n=1$ and $M=\Lambda$, becomes
\begin{equation}
\label{jeffmyers7}
C^{(1)}\sim V_{d-1}\Lambda^{d-1}
\end{equation} 
where $\Lambda$ is the cut-off and $M$ a parameter that characterizes the reference state. This result is in accordance with \eqref{jeffmyers3}.

In our approach, in order to get the proposed holographic complexity \eqref{jeffmyers3}, we assume that the period $\beta$ of the thermal circle is of the order of the lattice spacing $\delta$. Indeed, for $\beta$ very small and $t$ of the order of $\beta$ ($t\sim\beta$), the complexity becomes
\begin{equation}
C^{(1)}\sim V_{d-1}\beta^{-(d-1)}\nonumber
\end{equation}
and is similar to the proposed expression in ref. \eqref{jeffmyers3} with $\beta\sim\delta$.

\section*{Acknowledgments} 
This work was supported in part by the Natural Sciences and Engineering Research Council of Canada. We also thank Shira Chapman for her helpful comments and discussions.

\end{document}